\documentclass[11pt]{article}
\usepackage{mathptmx}
\usepackage[T1]{fontenc}
\usepackage{pifont}
\usepackage{amsmath}
\usepackage{graphicx}
\usepackage{amssymb}

\makeatletter
\@ifundefined{textcolor}{}
{%
 \definecolor{BLACK}{gray}{0}
 \definecolor{WHITE}{gray}{1}
 \definecolor{RED}{rgb}{1,0,0}
 \definecolor{GREEN}{rgb}{0,1,0}
 \definecolor{BLUE}{rgb}{0,0,1}
 \definecolor{CYAN}{cmyk}{1,0,0,0}
 \definecolor{MAGENTA}{cmyk}{0,1,0,0}
 \definecolor{YELLOW}{cmyk}{0,0,1,0}
 }

\newcommand{\C}[1]{{\mathcal{#1}}}
\newcommand{\pa}{\partial}

\begin{document}


\title{Submitted to ``Annual Review of Condensed Matter Physics''\\
\vskip 1.5cm Dynamics of Simple Cracks}
\author{Eran Bouchbinder$^1$,  Jay Fineberg$^2$, and M. Marder$^3$\\\\
\small{$^1$Department of Chemical Physics, Weizmann Institute of Science, Rehovot 76100, Israel}\\
\small{$^{2}$Racah Institute of Physics, Hebrew University of Jerusalem, Jerusalem 91904, Israel}\\
\small{$^{3}$Center for Nonlinear Dynamics and Department of Physics,}\\
\small{The University of Texas at Austin, Austin, Texas 78712, USA}}

\maketitle
\begin{abstract}
Cracks are the major vehicle for material failure, and often
exhibit rather complex dynamics. The laws that govern their motion
have remained an object of constant study for nearly a century.
The simplest kind of dynamic crack is a single crack that moves
along a straight line. We first briefly review current
understanding of this ``simple'' object. We then critically
examine the assumptions of the classic, scale-free, theory of
dynamic fracture, and note when it works and how it may fail if
certain of these assumptions are relaxed. A number of examples is
provided, where the introduction of physical scales into this
scale-free theory profoundly affects both a crack's structure and
the resulting dynamics.
\end{abstract}

\section{Introduction}

\subsection{Physics of Cracks}

The ability of solids to withstand mechanical forces is one of their
fundamental properties. Solids divide roughly into two classes, brittle
and ductile. The brittle solids break easily, catastrophically. Ductile
solids flow or bend before they break and are more resilient. There
is a simple thermodynamic definition of an ideally brittle solid.

Consider any solid sample of height $b$ whose top and bottom
boundaries are displaced by distances $\pm\delta/2$, as shown in
Fig. \ref{fig:setup}. If the total displacement $\delta$ is sufficiently
small, then the restoring force $F$ on the boundaries will be
$F=CA\delta/b,$ where $C$ is an elastic constant, and the change
in energy of the solid must be
\begin{equation}
\Delta E=\frac{1}{2}CA\delta^{2}/b.\label{eq:deltae}\end{equation}
For all solids one can also define the surface energy $\gamma,$
where $2\gamma$ is the minimum energy per area needed to separate
bonds and cut the solid into two pieces. For a given solid
$\gamma$ may depend upon the angle of the cut, but we will assume
for simplicity that the solid is oriented so that $\gamma$ has a
minimum along the plane perpendicular to $\vec{F.}$ Then there is
a critical displacement $\delta_{G}$ for which the solid will have
lower energy by separating into two pieces than by continuing to
withstand tension. This critical displacement is given by
\begin{eqnarray}
2\gamma A & = & \frac{1}{2}CA\delta_{G}^{2}/b\label{eq:deleqn}\\
\Rightarrow\delta_{G} & = & \sqrt{4\gamma b/C}\Rightarrow
F/A=\sqrt{4\gamma C/b}.\nonumber \end{eqnarray} and is called the
\emph{ Griffith point.} In the limit of large systems, where
$b\rightarrow\infty,$ the force per area goes to zero. This means
that the approximation of Eq. \ref{eq:deltae} is as accurate as
desired. Thus, there exists a universal relation
between energy, force, and boundary displacement shown in Fig.
\ref{fig:setup}, and given by \begin{equation} \Delta
E=\begin{cases}
\frac{1}{2}CA\delta^{2}/b & \mbox{ for }\delta<\delta_{G}\\
\frac{1}{2}CA\delta_{G}^{2}/b & \mbox{ for
}\delta\geq\delta_{G}.\end{cases}\ \ F=\begin{cases}
CA\delta/b & \mbox{ for }\delta<\delta_{G}\\
0 & \mbox{ for
}\delta\geq\delta_{G}\end{cases}\label{eq:universal}\end{equation}
 For large enough samples this relation is exact.

A solid that actually obeys Eq. \ref{eq:universal} is ideally
brittle. Ceramics such as silicon, and some glasses come close.
However, for most materials the whole picture is misleading or
simply wrong. It is not enough for thermodynamics to say that a
solid can lower its energy by dividing into two pieces. There must
actually exist some physically feasible way for the separation to
take place. Accomplishing this separation is the role of cracks.
Cracks are nonequilibrium propagating dissipative structures.
Initiating at a weak spot, they travel across solids under
mechanical stress and separate them into multiple pieces. The
questions of which solids are brittle, which are ductile, and how
much force or energy truly is needed in order to cause a solid to
fail come down to the questions of when cracks propagate and how.

\subsection{Scope of this article}

Two of us wrote a review 10 years ago that laid out the basic
mechanical theory of cracks and then described conditions under
which the tips became unstable \cite{Fineberg.99}. The opportunity
to write a second review raised a question of how to cover
additional topics in an article that would still be fairly
self-contained. We decided to focus on subjects that on the one
hand represent recent developments, but on the other hand are in
some ways more elementary than those we discussed before. We
examine both experimental and theoretical issues that arise when
one checks carefully if the theory of fracture is complete.

The fundamental theory of fracture, called Linear Elastic Fracture
Mechanics (LEFM), is now a well
established body of knowledge. The
mathematical foundations were carefully laid by investigators such
as Irwin \cite{Irwin.57}, Rice \cite{Rice.68} and Willis \cite{Willis.67}, and are
documented in textbooks such as those by Broberg
\cite{Broberg.99}, Freund \cite{Freund.90}, and Slepyan
\cite{Slepyan.02}. By itself, this theory of fracture does not
have a length scale within it, and it must be supplemented with
some physical information about what makes cracks start to
propagate that can either be obtained from experiments, or deduced
from additional calculations. Here we will discuss a number of
different cases where scale-free fractures interact with phenomena
involving specific physical scales.

In contrast with physical theories for solid properties such as
electrical conduction, physical theories of fracture are still in
a primitive state. The ability to predict material properties
based upon knowledge of atomic constituents or other underlying
properties, is limited. We have assembled recent advances in
describing detailed processes near crack tips in the hopes that,
eventually, the understanding of crack tip mechanics will reach
the point where crack direction and speed come under reliable
experimental control. It may then be possible to design materials
so that cracks begin to move at designated thresholds. At the
present time these goals are not in reach, as we describe in our
concluding section on conclusions and open questions.

\section{Fracture Mechanics}

\subsection*{Linear Elastic Fracture Mechanics\label{sub:Linear-Elastic-Fracture}}

\begin{figure}
\begin{centering}
\includegraphics[width=.9\textwidth]{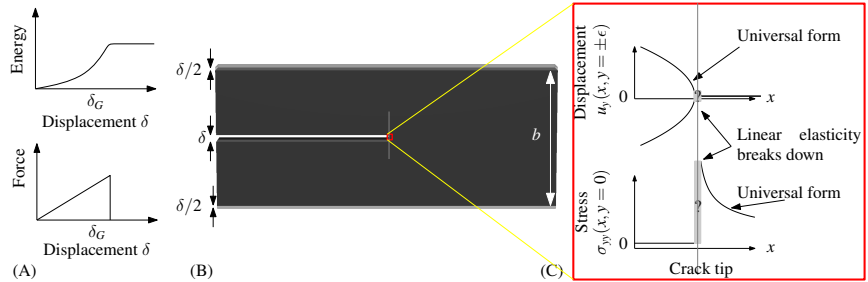}
\par\end{centering}

\caption{(A) Universal relation between energy, force and
displacement for ideally brittle solids in tension. The energy reaches a
plateau and force vanishes at the Griffith point $\delta_{G}$. (B)
Geometry of sample for which these universal relations between
energy, force, and displacement can be observed. The top and
bottom of the sample are gripped rigidly and moved apart by
distance $\delta/2$. A crack through the middle of the sample
provides a way for it to break in half. (C) Closeup view of
universal features of displacements and stresses near the tip of
the crack in (B). Very close to the crack tip linear elasticity
breaks down; sometimes it can be replaced by other theories
described later in this paper.\label{fig:setup}}

\end{figure}

When a crack exists in a material, any externally applied stress
undergoes a very large amplification at its tip. This was first
noted by Inglis \cite{Inglis.13}, who demonstrated amplification
of the stresses at the tip of an elliptical hole in an otherwise
uniformly stressed sheet. Irwin and Orowan
\cite{Orowan.34,Irwin.57} later showed that if a \emph{crack}
exists in a sheet assumed to obey linear elasticity, the stress
field $\sigma_{ij}$ actually becomes singular at its (infinitely
sharp) tip. The stress (see Fig. \ref{fig:setup}C) at a distance
$r$ from the tip takes the universal singular form
$\sigma_{ij}\propto K/\sqrt{r}$ where the coefficient $K$ of the
singularity is called the stress intensity factor. Depending on
the symmetry of loading, there can be three independent fracture modes
\cite{Freund.90,Broberg.99,Lawn.93}, Mode I - fracture in pure
tension, Mode II - fracture in pure in-plane shear, and Mode III -
fracture by out-of-plane or ``tearing''. The stress field due to
each of these modes has the same universal form, and there may be,
in general, three independent stress intensity factors. Here, we
will only consider fracture under pure tension (Mode I).

The existence of a singular stress is, itself, insufficient to
make a crack propagate, since the singularity is always cut off at
least at the atomic scale, where it may or may not be large enough
to snap bonds. A better way to determine if cracks can move is to
note that new surface must be created and this requires energy.
Defining the energy required to create a unit area as the
``Fracture Energy'', $\Gamma$, a simple generalization of
Griffith's idea \cite{Griffith.20} suggests that cracks begin to
move when the potential elastic energy per unit area released by a
unit extension of a crack becomes equal to $\Gamma$ (cf. Eq.
\ref{eq:deleqn}). Even if this condition of energy balance
predicts the onset of motion, once a crack starts to move the
kinetic energy due to material motion around the moving crack has
to be taken into account. Accounting for this energy led to the
first equation of motion for a crack, which Mott initially
obtained by dimensional analysis \cite{Mott.47,Dulaney.60}. This
equation predicted that in a two-dimensional medium of infinite
extent, a crack should continuously accelerate as a function of
its instantaneous length to a limiting, but finite, asymptotic
velocity. This speed limit for a crack is due to physical constraints
on energy transport. The crack tip requires energy to
extend. If this energy has to be transported to the tip from
remote parts of the medium by elastic waves, the speed of a crack
must certainly be limited by the propagation speed of those waves.
This asymptotic limit can be reached by either increasing the
amount of energy driving the crack or by reducing the fracture energy $\Gamma$ to
zero. Stroh \cite{Stroh.57} noted that for $\Gamma\rightarrow0$, a
crack propagating at its asymptotic velocity is equivalent to a
disturbance moving along a free surface and therefore predicted
the crack's limiting velocity to be the Rayleigh wave speed,
$V_{R}$, which is the highest speed that a wave can move along a
free surface.

These intuitive ideas were later
\cite{Eshelby.70,Kostrov.74,Freund.90,Willis.67,Rice.68} shown to
be rigorously correct, when a quantitative theory of dynamic
fracture was developed. This complete theory is called Linear
Elastic Fracture Mechanics (LEFM). Assuming that the medium always
obeys linear elasticity, the theory predicts that the general form
for the stress singularity at the tip of a crack as a function of
the distance $r$ and angle $\theta$ from the tip of a crack moving
at velocity $v$ is given by \cite{Freund.90}

\begin{equation}
\sigma_{ij}=\frac{K}{\sqrt{r}}f_{ij}(v,\theta),\label{eq:stress_field}
\end{equation}
where $f_{ij}(v,\theta)$ is a known universal function and $K$ can
be explicitly calculated for any externally applied loads. $K$ has the dimensions of stress $\times$ length$^{1/2}$, and thus must depend on a macroscopic geometric lengthscale. For
sufficiently large samples, there is always a region around the
crack tip where the singular term in Eq. \ref{eq:stress_field}
dominates any other contribution to the stress field. This fact
can then be used to find the energy release rate, $G$, defined as
the energy per unit extension per unit width  that is flowing into the tip
of a crack. $G$ is related to $K$ by \cite{Freund.90}:
$G=(1-\upsilon^{2})K^{2}A(v)/E,$ where $\upsilon$ and $E$ are,
respectively, the Poisson ratio and Young's modulus of the medium.
$A(v)$ is a universal function of the instantaneous crack
velocity $v$. With this expression in hand, the general equation
of motion for a crack can be obtained by using the energy balance
criterion, equating $G$ to $\Gamma$. Energy balance, however,
needs to be supplemented by an additional condition that tells us
in what\emph{ direction} a crack will move. In general, no such
first-principles condition exists. One commonly used assumption,
called the ``principle of local symmetry''
\cite{Goldstein.Salganik,Cotterell.Rice}, tells us that a crack
will locally align itself in a direction so as to negate any shear
component at its tip. For quasistatic (``slow'') fracture this
idea has been shown to be justified by symmetry considerations
\cite{Hodgdon.93,Oleaga.01} and has been used, for example, to
quantitatively describe
\cite{Marder.94.pre,Sasa.94,Adda-Bedia.95,Bahr.95,Sumi2000,Yang2001,Bouchbinder.03,Pham.08,Corson.09}
experimental measurements of slow oscillating cracks in heated
strips \cite{Yuse.93,Ronsin.95}. Other closely related path selection criteria were proposed and tested, see for example \cite{Pollard.93}. Although the ``principle of local symmetry'' is often assumed for rapid
cracks, there is, however, neither fundamental or experimental
justification for this (or any other) crack path selection
criterion.

An additional tacit assumption is also used; that all of the
dissipation in the system occurs within a  ``small'' region
surrounding the tip of the crack. This means that all of the
complex dissipative mechanisms, inherent in the fracture process,
have to occur in a region for which the surrounding stress field
is entirely dominated by the singular stress field described by
Eq. \ref{eq:stress_field}. This condition, called the condition of
small-scale yielding, is, in essence, the definition of a brittle
material - which extends the concept of an ideal brittle material
(cf. Fig. \ref{fig:setup}A). The small zone surrounding the tip in
which all nonlinear and dissipative processes occur is called the
process zone. Thus, $\Gamma$ need not necessarily be the energy
cost to break a single plane of molecular bonds \textendash{} it
embodies all of the complex nonlinear processes that are driven by
the putatively singular stress field near the crack tip.
Small-scale yielding is one of the foundations of fracture
mechanics, as it ensures the equivalence of the global energy
balance criterion used by Mott with the locally formulated
equation of motion: \begin{equation}
G=\Gamma=(1-\upsilon^{2})K^{2}(l,v)A(v)/E.\label{eq:eqn_motion}\end{equation}
 Freund \cite{Freund.90} demonstrated that for straight semi-infinite
cracks in infinite plates, under a variety of loading conditions
\cite{Freund.90}, the dynamic stress intensity factor $K(l,v)$ has
a separable form $K(l,v)=K(l)k(v)$, where $K(l)$, which can be
explicitly calculated, is solely dependent on the external loading
configuration and the instantaneous value of the crack length,
$l$. Both $k(v)$ and $A(v)$ are universal functions of only $v$.
Freund \cite{Freund.90} showed that, to a good approximation
$k(v)A(v)\approx1-v/V_{R}$ yielding
\begin{equation}
\Gamma=G\approx\frac{(1-\ensuremath{\upsilon^{2}})K^{2}(l)}{E}\cdot(1-v/V_{R})\label{eq:Gamma}\end{equation}
Inverting Eq. (\ref{eq:Gamma}) yields the following prediction for
an equation of motion:

\begin{equation}
v=V_{R}\left[1-\frac{\Gamma
E}{(1-\upsilon^{2})K^{2}(l)}\right]\label{eq:eqn_motion_2}\end{equation}
which is identical to the equation obtained earlier by dimensional
analysis for a crack in an infinite plate.

\begin{figure}
\begin{centering}
\includegraphics[width=0.9\textwidth]{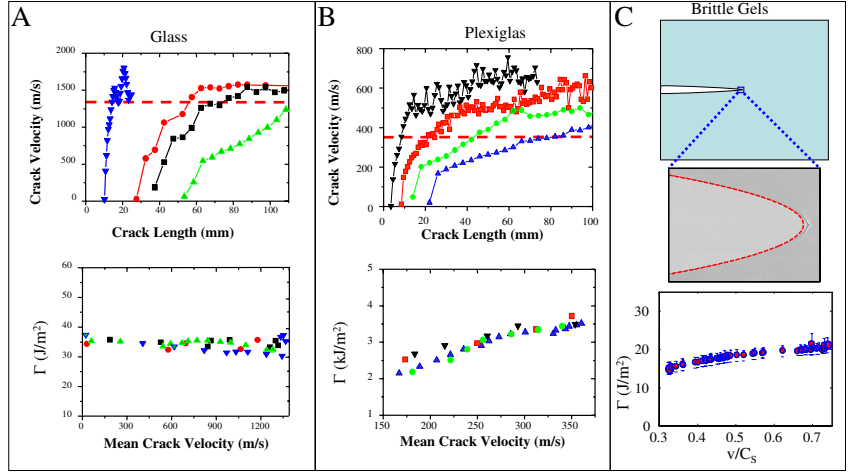}
\par\end{centering}

\caption{Fracture energy, $\Gamma$, derived using Eq.
\ref{eq:Gamma}, as a function of the mean velocity for soda-lime
glass (A) and PMMA (B). Top: Velocities of cracks driven with
different initial stresses and initial crack lengths as a function
of their instantaneous lengths. The data were smoothed over a 1mm
length, to filter out velocity fluctuations. The dashed lines
indicate the highest values of $v$ used to derive the
corresponding $\Gamma(v)$ (lower plots). These are 350 m/s (0.38
$V_{R}$) in PMMA and 1325 m/s (0.40 $V_{R}$) in glass. The
observed data collapse to a single (material-dependent) function,
$\Gamma(v)$ is a quantitative validation of the equation of motion
derived for a single crack. Data were taken from \cite{Sharon.99}.
Values of $V_{R}$ for PMMA and soda-lime glass are, respectively,
930m/s and 3,300m/s. (C) $\Gamma(v)$, derived by using the
parabolic profile of the crack tip predicted by LEFM. These
measurements were made possible by \cite{Bouchbinder.08a,Livne.08}
using soft brittle gels, where $V_{R}$ =
5.3m/s.\label{fig:test_LEFM-1}}

\end{figure}

Does this equation of motion work? Sharon et al. \cite{Sharon.99}
conducted a quantitative test of Eq. \ref{eq:eqn_motion_2} where
the detailed dynamics of rapidly propagating cracks were measured
in both PMMA (Plexiglas) and soda-lime glass, two brittle
amorphous materials. The experiments measured the instantaneous
values of $l$ and $v$ under loading conditions whose geometry
enabled a precise calculation of $K(l)$. The values of $\Gamma$
were derived using Eq. \ref{eq:Gamma} for widely different
experimental conditions. $\Gamma$ is a material property, but need
not be velocity independent and, indeed, is generally a function
of $v$. If Eq. \ref{eq:eqn_motion_2} is valid, the values of
$\Gamma(v)$ derived from Eq. \ref{eq:Gamma} using measurements of
$l$ and $v$, should all collapse onto a single curve for each
material. The results of these experiments were surprising. As we
see in Fig. \ref{fig:test_LEFM-1}, the experiments for both
materials indeed collapse onto single curves for both glass and
PMMA at crack velocities that are less than about $0.4V_{R}$ in
each material. Thus, Eq. \ref{eq:eqn_motion_2} is indeed in
excellent quantitative agreement for $v<0.4V_{R}$ for both glass
and PMMA, two materials whose micro-structure is entirely
different. These results agree with earlier results of Bergqvist
\cite{Bergkvist.74} and Ravi-Chandar et al \cite{Ravi-Chandar.84a}
that showed that cracks appear to behave in a way that is
consistent with LEFM, for relatively low crack velocities.

\begin{figure}[h]
\begin{centering}
\includegraphics[width=0.9\textwidth]{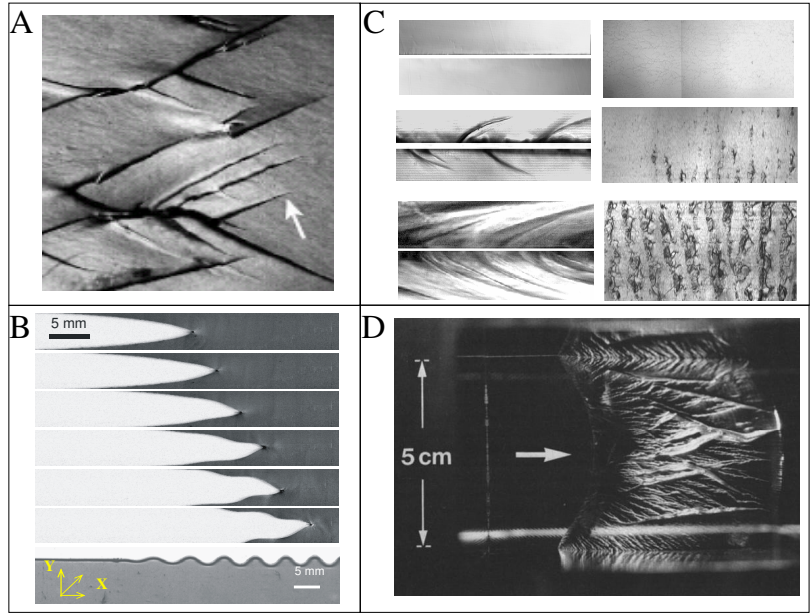}
\par\end{centering}

\caption{(A) A cross-hatching instability in slow hydrogel Mode I
fracture due to straight macrosteps on the crack surfaces
\cite{Baumberger.08}. (B) A high-speed crack tip oscillatory
instability in Mode I fracture of an elastomer gel
\cite{Livne.07}. (C) A side view of the side-branching instability
in rapid Mode I fracture in Plexiglas (left column) and the
accompanying crack surfaces (right column)
\cite{Fineberg1997,Fineberg.99}. (D) Complex features on the
fracture surfaces of a continuously twisting mixed Mode I+III
crack in Plexiglas \cite{Cooke.96}.\label{Fig:complex-1}}

\end{figure}

What happens when $v>0.4V_{R}$? At a critical velocity,
$v_{c}\sim0.4V_{R}$ a single crack state become unstable
\cite{Fineberg.91,Fineberg.99,Sharon.95} to one in which the main
crack is continuously sprouting frustrated microscopic branched
cracks, as shown in Fig. \ref{Fig:complex-1}C. This
``micro-branching'' instability causes rapid oscillations in the
instantaneous crack velocity and the formation of non-trivial
structure on the fracture surface. There is currently no
first-principles theory for the origin of this and other
instabilities, although many serious attempts have been made
 \cite{Marder.93.prl,Gao.93,Abraham.94,Ching.94,Marder.95.jmps,Ching.95,Gao.96,Adda-Bedia.Ben-Amar.96,Ching.96a,Ching.96b,Ching.96c,Gumbsch.97,Brener.98,Sander.99,Adda-Bedia.99,Cramer.00,Boudet.00,Pla.00,Heizler.02,Buehler.03,Bouchbinder.04,Adda-Bedia.04,Bouchbinder.05a,Bouchbinder.05b,Buehler2006,Pilipenko.07,Bouchbinder.07}.
It seems clear, however, that the key does not lie within the
framework of LEFM. In fact, once a simple single-crack state is
lost, we have little fundamental understanding of fracture
dynamics. Some examples of the complexities that such non-trivial
states can produce are shown in Fig. \ref{Fig:complex-1}.

Does LEFM break down for $v>0.4V_{R}$ if instabilities are
suppressed? Recent experiments
\cite{Bouchbinder.08a,Livne.07,Livne.08,Bouchbinder.09} show us
that, as long as its underlying assumptions are met, LEFM is
entirely accurate. For example, the displacement fields
surrounding a crack's tip can be obtained by integrating Eq.
\ref{eq:stress_field}. In particular, displacements normal to the
crack's propagation direction ($\theta=0$), predicts a parabolic
crack tip opening
$u_{y}(r<0,\theta=\mbox{\ensuremath{\pi}})\propto
K(l,v)\cdot\sqrt{r}$ behind the crack (Fig.
\ref{fig:setup}C-top), whose curvature is determined by $K(l,v)$.
Crack instabilities have been suppressed to up to $v\sim0.9V_{R}$ in
soft brittle gels \cite{Livne.07}. In these materials, $V_{R}$ is
1000 times lower than in standard materials. This enabled direct
measurements of the crack tip profile of rapidly propagating
cracks \cite{Livne.08}. These experiments show that, as long as we
are not too close to the crack tip, LEFM provides excellent
quantitative agreement with measurements. Values of $\Gamma(v)$
collapse onto a well-defined function over a very large range of
velocities, as Fig. \ref{fig:test_LEFM-1}C shows, when calculated
using values of $K(l,v)$ obtained by measuring the crack tip
curvature.

\subsection{LEFM - when does it work?}

Sometimes LEFM provides an excellent quantitative description of both
a crack's motion and of the elastic fields surrounding its tip. This
happens whenever the, at times subtle, fundamental assumptions underlying
LEFM are valid. Let us list these assumptions:
\begin{itemize}
\item Only the dynamic behavior of straight single-crack states are accounted
for. Once more complex dynamic states (e.g. micro-branching) occur,
LEFM fails to accurately predict their behavior. In fact, there is
currently no first-principles criterion for what path a crack must
take.
\item LEFM assumes that linear elasticity is valid away from the crack tip.
In principle, for large enough samples, this assumption can always
be made as accurate as desired, but in practice for samples of fixed
size it must be checked.
\item Small-scale yielding is observed $\Leftrightarrow$ all nonlinear
and/or dissipative processes are assumed to occur in a region of
negligible size at a crack's tip. As LEFM is a scale-free theory,
this tacitly means that the scales at which these complex
processes occur are negligible. We will see that this tacit
assumption is not necessarily justified; the existence of
additional scales within the crack tip region can and does have a
large effect on crack dynamics.
\item A tacit assumption concerning ``energy balance'' is that energy
has to flow \emph{ into} a crack's tip, otherwise a crack cannot
propagate.
\end{itemize}
In the current ``state of the art'' in fracture we have little
fundamental understanding of the complexities of a crack's motion
once a crack decides to stray from a straight path or become
unstable. These questions are hard ones, and although there is a
number of possible directions to take, we have no clear answers.
In the following sections we have, instead, chosen to describe
what we currently know about ``simple'' straight cracks. We will
do this by describing a number of interesting cases where  theory and experiment
move beyond limits constraining LEFM. By doing this we
hope to provide both insights into the character of the ``simple''
crack state, and to, perhaps, provide a springboard for
understanding more complex fracture dynamics.

\section{Tales of Scales}

\subsection{The Importance of Small Scales near a Crack's Tip}

LEFM is a scale-free theory. Scales, however, must be introduced
(either tacitly or explicitly) to obtain a closed theoretical
description of a crack's dynamics. A necessary scale, for example,
is inherent in the idea of the process zone, which both cuts off
the singular fields given by Eq. \ref{eq:stress_field} and
provides a scale below which dissipation and material nonlinearity
take place. A central question in fracture is whether the details
of these small scales are important. Is a proper description of
these scales needed to describe how a crack behaves, or do the
universal properties of fracture indicate that the details of the
fracture processes that take place at these scales are irrelevant
and simply serve to determine a material's fracture energy.

There is a number of hints that the character of these small
scales plays a major role in fracture dynamics. A classic way to
regularize the continuum singularity at the tips of cracks is to
introduce \emph{cohesive zones}, fictitious forces pulling the
crack's faces together behind the tip in just the right way so
that the singularity of the stress field
disappears\cite{Dugdale.60,Barenblatt.59;b}. In cohesive
zone models, these fictitious forces acquire equations of motion
so that they can travel along with the crack, cancelling out the
singularity as they go. Results of working out such models are
disquieting. It is difficult to obtain stable straight-ahead
propagation of a single crack at \emph{any} velocity
\cite{Ching.96b,Langer.98}. As noted by Lobkovsky and Langer
\cite{Langer.98}, this entire class of models {}``seem[s] to be
highly sensitive to details that ought to be physically
unimportant''.

This point of view is reinforced by numerical simulations. For
example, finite element calculations of dynamic cracks
\cite{Miller.99} with cohesive zones can provide quantitative
agreement with experimental descriptions of both stable cracks and
cracks undergoing the micro-branching instability described in
Fig. \ref{Fig:complex-1}C. However, it turned out that even the
existence of micro-branching in numerics depends on the mesh size
used in the simulations, and the experimental phenomena disappear
in the limit where the mesh size goes to zero
 \cite{Falk.01}.

In this section, we will first describe three approaches in which
a well-defined length scale at the crack's tip occurs naturally.
Two of these are continuum theories in which scales are formed
either by (1) relaxing the assumption that materials are elastic
up until the point where dissipation occurs or (2) the
introduction of a phenomenological scalar field that represents
the material damage that occurs around the tip. Both of these
approaches define a new dynamic scale at the tip. A third class of
models will be described in which the medium is an ideal crystal,
which therefore possesses an intrinsic length-scale, the lattice
size. We will see that in all of these approaches the existence of
a length scale at the crack tip has important consequences on both
the dynamics and form of propagating cracks.

We then show that an entire new class of single-crack solutions
can arise when the Griffith condition is supplanted by the
introduction of new energy scale. These solutions, which are
wholly physical, surpass the speed limit set by LEFM by violating its
assumption about how energy is conserved near a crack tip.

\subsection{\label{sub:Small-Scales-Dominated}Small Scales Dominated by Nonlinear
Elasticity}

An essential feature of crack propagation is the high
concentration of deformation in the immediate vicinity of a
crack's tip, as given by LEFM's $\sqrt{r}$-singular fields. LEFM,
however, is inherently confined to small material deformations.
Therefore, any attempt to understand the physics of the process
zone involves first the question of how and where LEFM breaks
down near the tip of a crack. To address this question, consider
the interatomic potential by which particles interact. Adopt now
the simplified assumption that the material's response until the
point of failure is completely contained within this interaction
potential (i.e. neglect features such as dislocation motion, or
changes in the underlying geometry of amorphous solids). Linear
elasticity is described by the harmonic approximation about the
equilibrium state at the bottom of the potential well. Plastic
deformation, damage evolution and finally fracture correspond to
interatomic separations far from the minimum of the potential,
where the interaction energy decreases significantly. It is both
natural and physically sound to assume that in numerous materials
before one of these irreversible processes occurs, the harmonic
approximation will first break down via \emph{reversible}
nonlinear deformation. Therefore, the \emph{first} physical
process that intervenes when LEFM breaks down should result from
corrections that arise from nonlinear elasticity. The first
generic nonlinear elastic contribution would be expected to come
from quadratic corrections to linear elasticity.

To model this behavior at a macroscopic level, we expand a general
stress vs. displacement-gradient relation up to second order \begin{equation}
\sigma\simeq\mu\pa u-\bar{\mu}(\pa u)^{2}+\C O[(\pa u)^{3}].\label{nonlinear_scalar}\end{equation}
 For simplicity, we suppress the tensorial nature of the quantities
involved and let $\sigma$ and $\pa u$ schematically denote stress
and displacement-gradient (strain) measures, respectively ($u$
denotes a displacement and $\pa$ schematically denotes a spatial
derivative); $\mu$ and $\bar{\mu}$ denote the first and second
order elastic moduli, respectively. We term the theory that is
based on this expansion ``weakly nonlinear fracture mechanics''
\cite{Bouchbinder.08a,Bouchbinder.08b,Bouchbinder.09}, as higher
order nonlinearities are neglected in Eq.
\ref{nonlinear_scalar}. To explore the implications of Eq.
\ref{nonlinear_scalar} on fracture dynamics, we schematically
write the momentum balance equation as
\begin{equation}
\pa\sigma=0\ ,\label{EOM_scalar}\end{equation}
where, for simplicity, we omit the inertial term on the right hand
side. We then expand the displacement $u$ in powers of the magnitude
of the (small) displacement-gradient $\epsilon\simeq|\pa u|$ and
to second order obtain \begin{equation}
u\simeq\epsilon\tilde{u}^{(1)}+\epsilon^{2}\tilde{u}^{(2)}\equiv u^{(1)}+u^{(2)}\ .\label{expansion_scalar}\end{equation}
 Substituting Eq. \ref{expansion_scalar} in Eq. \ref{EOM_scalar}
and expanding to first order, we obtain
\begin{equation}
\mu\,\pa^{2}u^{(1)}=0\ ,\label{EOM_scalar_1st}
\end{equation}
which is our schematic analog of the Lamé equation \cite{Freund.90}.
LEFM tells us that this equation, together with the traction-free
boundary conditions on the crack faces, leads to the following
asymptotic behavior near the tip of a crack \cite{Freund.90}
\begin{equation} \pa u^{(1)}\sim\frac{K}{\mu\sqrt{r}}\
,\label{scalar_1st}\end{equation}
 where $K$ is the stress intensity factor and $r$ is measured from
the crack tip, cf. Eq. \ref{eq:stress_field}. In terms of the displacement-gradients Eq. \ref{EOM_scalar},
to second order, becomes \begin{equation}
\mu\,\pa^{2}u^{(2)}-\bar{\mu}\,\pa(\pa u^{(1)})^{2}=0\ .\label{EOM_scalar_2nd}\end{equation}
 This is analogous to a Lamé equation for $u^{(2)}$ with an effective
body-force which scales as \begin{equation}
\bar{\mu}\,\pa(\pa u^{(1)})^{2}\sim\frac{\bar{\mu}K^{2}}{\mu^{2}r^{2}}\ .\label{body_force}\end{equation}
 By inspection, Eq. (\ref{EOM_scalar_2nd}) admits a solution of the
form \begin{equation}
\pa u^{(2)}\sim\frac{\bar{\mu}K^{2}}{\mu^{3}r}\ .\label{scalar_2nd}\end{equation}
 We therefore see that simply accounting for quadratic contributions
to material stress-strain relations already leads to the
interesting conclusion that the first nonlinear correction to the
asymptotic LEFM fields is characterized by a $1/r$
displacement-gradient singularity. The simple scaling-like
considerations described above capture the essence of the
solution, although the complete dynamic (i.e. with inertial
effects) weakly nonlinear solution is, of course, much more
mathematically involved \cite{Bouchbinder.08a,Bouchbinder.09}. For
example, the complete solution, which depends on the tip polar
coordinates $r$ and $\theta$, shows that $1/r$
displacement-gradients terms arise from both $r$-independent and
$\log{(r)}$ displacement contributions; the latter has an
important implication as it yields significant distortion of the
parabolic crack profile when the tip is approached.

\begin{figure}
\begin{centering}
\includegraphics[width=0.6\textwidth]{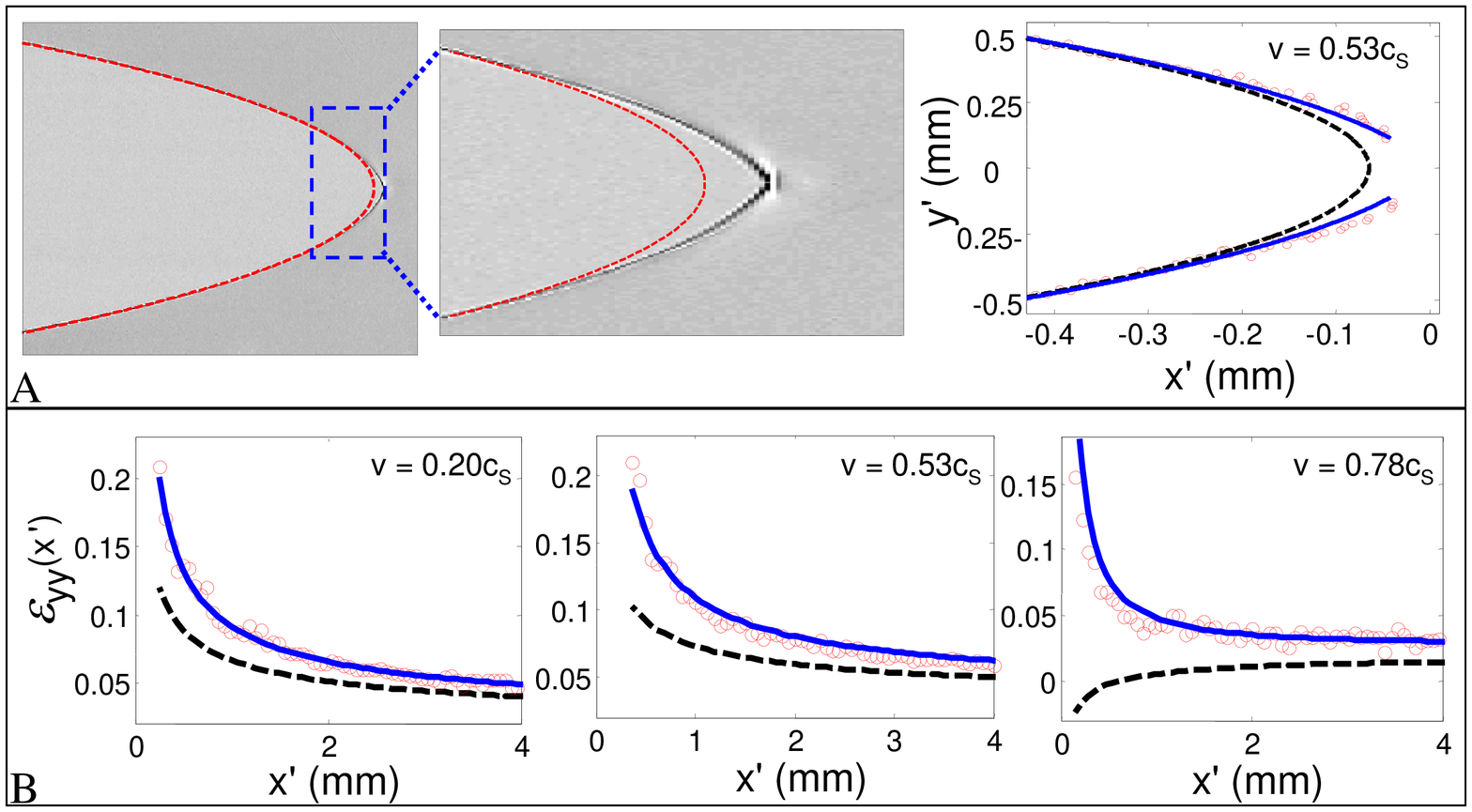}
\par\end{centering}

\caption{(A) Left: An experimental picture of a crack, with a
parabolic LEFM fit, cf. Fig. \ref{fig:test_LEFM-1}. Middle: A zoom
in on the crack tip. Right: A comparison of the weakly nonlinear
theory (solid line) with the measured crack tip opening profile
(circles). The LEFM parabolic fit is added (dashed line).
$(x',y')$ are the coordinates in the deformed (laboratory) frame.
(A) A comparison of the weakly nonlinear theory (solid line) with
the measured strain $\varepsilon_{yy}$ along the symmetry line
ahead of the crack tip, for three crack velocities. The LEFM
predictions are added (dashed lines). See text for more
details.\label{fig:NonLinear}}

\label{fig}
\end{figure}

Having addressed the question of \emph{how} LEFM breaks down under the stated assumptions, we
proceed now to consider the question of \emph{where} this happens
in space. According to the preceding discussion, LEFM breaks down
when the second order correction $\pa u^{(2)}$ becomes
non-negligible with respect to $\pa u^{(1)}$. A plausible estimate
for this condition is obtained from the condition that the second
order nonlinear contribution to the stress-strain relation becomes
larger than 10\% of the leading, linear, term. Thus,
\begin{equation} \frac{\pa u^{(2)}}{\pa u^{(1)}}\simeq 0.1\ .\label{scale_nonlinear}\end{equation}
 This immediately implies that higher order corrections like $1/r^{3/2}$ are negligibly
small in this region. As the strains surrounding any crack tip are
approaching a mathematical singularity, this condition must always
be satisfied at some scale. Denoting the scale at which Eq. \ref{scale_nonlinear}
is first satisfied by $\ell_{nl}$ and using Eqs. \ref{scalar_1st}
and \ref{scalar_2nd}, we obtain
\begin{equation}
\ell_{nl}\sim\frac{\bar{\mu}^{2}\, K^{2}}{\mu^{4}\,0.1^{2}}\ .\label{scale_nonlinearA}
\end{equation}
 This simple analysis reveals the emergence of a new length-scale associated
with weak elastic nonlinearities. It is important to note that
$\ell_{nl}$ is \emph{dynamic} as its value increases with the
energy driving the crack $\sim K^{2}/\mu$ or, equivalently, with a
crack's velocity.

Is the scale $\ell_{nl}$ in Eq. \ref{scale_nonlinear} physically
relevant? Nonlinear elasticity in large-scale measurements is
often not readily apparent. Brittle materials nearly always appear
linear elastic until the onset of irreversible deformation. These
measurements can be, however, quite deceiving. Failure in these
tests is \emph{not} due to intrinsic material behavior, but
generally takes place via the activation of cracks or defects that
are already in place within the material. On the other hand,
cracks generally propagate into locally virgin material. The
deformations near the tip of a crack are large enough to be well
within the nonlinear elastic regime. As defects are ubiquitous in
most macroscopic materials, these effects may be nearly
undetectable in large-scale measurements. This does not mean,
however, that they are either unimportant or do not commonly
occur.

With this in mind, irreversible (plastic) deformation may commonly
be preceded by elastic nonlinearities as the displacement-gradients
increase. Thus, the plastic deformation scale $\ell_{pl}$ is expected to satisfy
\begin{equation}
\ell_{nl}\gtrsim\ell_{pl}\ .\label{plastic_scale}\end{equation}
 For strongly nonlinear elastic materials (gels, rubber etc.) we expect
a scale separation of the form $\ell_{nl}\!\gg\!\ell_{pl}$, while
for other materials we cannot rule out the possibility that $\ell_{nl}$
is not very different from $\ell_{pl}$. In the former case, the
small scales near a crack's tip are dominated by nonlinear elasticity,
while in the latter case, dissipation may play a more decisive role.

The complete solution of the weakly nonlinear equations, which
generalizes the simple considerations above, was obtained in
Refs.~\cite{Bouchbinder.08a,Bouchbinder.09}. Can these solutions be
observed in experiments? Generally, this is a near-impossible task
since cracks propagate at near sound-speed velocities with (at
best) $\mu$m-scale crack tip openings. In addition, the second
order elastic constants are generally not well known, since in
macroscopic stress-strain tests of brittle materials failure, due
to the propagation of intrinsic cracks, occurs well before
sufficient strains are reached. Recent experiments, however,
circumvented these problems by utilizing brittle elastomers (gels)
\cite{Livne.08}. While the characteristic fracture phenomenology
of these soft materials is \emph{identical} to that of other
brittle amorphous materials \cite{Livne.05}, their sound speeds
are 1000 times lower. Furthermore, as these materials are very
compliant, but tough, they undergo large strains prior to
fracture. This allows precise measurements of their second-order
elastic constants. These properties enabled the first direct
experimental measurements of the deformation near the tip of rapid
cracks. A summary of the comparison of these measurements with the
explicit predictions of the weakly nonlinear theory is presented
in Fig. \ref{fig}. On the left hand side of the upper panel the
LEFM parabolic crack tip profile is shown, cf. the discussion of
this point in Sect. \ref{sub:Linear-Elastic-Fracture} and Fig.
\ref{fig:test_LEFM-1}. In the middle of the upper panel a zoom in
on the crack tip region is shown, where marked deviations from the
LEFM parabolic profile are observed. In \cite{Livne.08} it was
shown explicitly that this deviation is due to elastic deformation
and not due to an irreversible one.

On the right hand side of the upper panel we show the predictions
of the complete weakly nonlinear solution for the tip profile. We
observe that it agrees well with the LEFM parabolic form at large
scales (i.e. small deformation), but both qualitatively and
quantitatively (with \emph{no} adjustable parameters, since all
elastic properties can be measured in this material) captures the
deviation from it on smaller scales (below $\sim200\mu$m in this
example). The deviation is a direct consequence of the $\log{(r)}$
contribution to $u_{x}$ and directly demonstrates the existence of
the new terms predicted by the weakly nonlinear theory.

In the lower panel of Fig. \ref{fig} we present a comparison
between the weakly nonlinear theory and the measured deformation
\emph{ahead} of the crack tip. We focus on
$\varepsilon_{yy}\!=\!\pa_{y}u_{y}$ along the symmetry line
($\theta=0$) ahead of the crack for three propagation velocities,
including one that approaches the limiting speed, $V_R$. For the
two lowest velocities we observe marked deviations from the
$1/\sqrt{r}$ fields of LEFM, deviations that are accurately
captured by the weakly nonlinear theory (cf.
\cite{Bouchbinder.08a} for more details). These results again
explicitly demonstrate the validity of the weakly nonlinear theory
with its $1/r$ singularity. Moreover, the scale $\ell_{nl}$ in
which LEFM breaks down via elastic nonlinearities is captured by
the theory. Its value, which is in the mm range, can be easily
read off from the figure.

On the right hand side of the left panel we present the same
comparison for a crack propagating at $v\!=\!0.78c_{s}$. At this
very high velocity, higher than second order nonlinearities are
needed to accurately describe the data. This comparison, however,
highlights an important point. According to LEFM, a velocity
exists ($0.73c_{s}$ for an incompressible material) where
$\varepsilon_{yy}$ changes sign from positive to negative. In this
range of velocities $\varepsilon_{yy}$ predicted by LEFM is
small (due to kinematic functions
\cite{Freund.90,Livne.08,Bouchbinder.08a} that enter as
coefficients in the exact analog of Eq. \ref{scalar_1st}, but do
not appear in the simple derivation used above). As a result, in
this range of velocities the second order term in
$\varepsilon_{yy}$ becomes the \emph{dominant} contribution. Due
to this contribution, as seen in the figure, no change in sign
occurs. Thus, the weakly nonlinear theory retains our basic
physical intuition about fracture; material points straddling the
symmetry line must be separated from one another
($\varepsilon_{yy}\!>\!0$) to precipitate fracture. This contrasts
with the LEFM prediction in this range of velocities where
$\varepsilon_{yy}\!<\!0$ (cf. dashed line in the figure).

In summary, the weakly nonlinear theory of dynamic fracture
describes how LEFM breaks down in materials where the scale
defined by elastic nonlinearity is larger than the dissipative
scale. In addition to accurately describing detailed measurements
of the deformation near tips of very rapid cracks
\cite{Bouchbinder.08a,Bouchbinder.09}, this theory may also have
further profound implications for understanding crack tip
instabilities \cite{Fineberg.99,Bouchbinder.08a}. This theory, by
itself, does not account for either near-tip dissipation or
directly offer crack path predictions. The length-scale
$\ell_{nl}$ for the breakdown of LEFM may, however, play a central
role in explaining symmetry breaking instabilities. There are some
tantalizing hints that crack tip instabilities may occur at
approximately the length-scale $\ell_{nl}$ that emerges in this
theory. For example, $\ell_{nl}$ at high velocities correlates
well with the geometry-independent wavelength of crack path
oscillations discussed in \cite{Livne.07,Bouchbinder.07}. This
promising line of investigation should be further explored.

\subsection{\label{sub:Phase-Field-Models}Phase Field Models}

LEFM accurately predicts the transport of linear elastic energy
from the large scales, where external forces are applied, to the
small scales near the tip of a crack (the process zone), where
energy is dissipated in fracture. It is assumed that these scales
are coupled solely through the $\sqrt{r}$ -- singular field. This
approach, however, is inherently deficient in at least three
important aspects:
\begin{itemize}
\item The crack tip velocity $v$ cannot be determined self-consistently
within LEFM, as energy dissipation within the nonlinear near-tip region
is not described in this framework.
\item The path selected by a crack's tip remains undetermined in the framework
of LEFM unless supplemented by a path selection criterion (e.g. the
principle of local symmetry). Hence crack tip instabilities cannot
be explained by solely LEFM.
\item LEFM is a scale-free theory and hence any phenomenon that involves
a non-geometric length-scale is beyond its scope (e.g. some crack
tip instabilities).
\end{itemize}
All of these difficulties are directly related to the need to
account for the physics at the small scales near the tip of a
crack, where LEFM breaks down. A comprehensive theoretical
framework should explain how LEFM breaks down near the tip of a
crack, predict crack tip shape and velocity selection, quantify
energy dissipation near the tip of a crack and account for the
path selected by cracks. Crack tip instabilities are expected to
naturally emerge from such a theory.

A class of phenomenological phase field models has recently been
developed to approach this problem from a continuum perspective
\cite{Francfort.98,Aranson.00,Karma.01,Karma2004,Henry2004,Marconi.05,Spatschek.06,Henry.08,Corson.09,Hakim.09}.
In this approach, the fracture of materials is described as the
gradual accumulation of damage near the tip of a crack. Damage is
quantified mathematically by a scalar ``order parameter'' or
``phase field'' $\phi(\mathbf{x},t)$, where $\phi\!=\!1$
corresponds to an intact material and $\phi\,=\,0$ to a fully
broken one. The physical interpretation of the phase field $\phi$
as quantifying local damage is, in fact, not necessary and most
probably will not be advocated by all practitioners of this
approach. The phase field can be viewed as a mathematical
interpolation between LEFM and material failure, where
$\phi\!=\!1$ is favored when the linear elastic strain energy
density $\mathcal{E}_{el}$ is smaller than some threshold
$\mathcal{E}_{c}$, $\mathcal{E}_{el}\!<\!\mathcal{E}_{c}$, while
$\phi\!=\!0$ is favored when $\mathcal{E}\!>\!\mathcal{E}_{c}$. A
similar approach proved to be very fruitful in the context of
non-equilibrium crystal growth, when used to predict
solidification micro-structures.

The mathematical formulation of these ideas starts with the following
expression for the potential energy density as a function of $\phi$
and the displacement field $\mathbf{u}$ \begin{equation}
\mathcal{E}=\frac{\kappa}{2}|\nabla\phi|^{2}+g(\phi)\left(\mathcal{E}_{el}-\mathcal{E}_{c}\right)\ .\label{potentialE}\end{equation}
 $\mathcal{E}_{el}$ is the linear elastic energy density, which is
given in terms of linear elastic strain tensor $\boldsymbol{\varepsilon}$,
as \begin{eqnarray}
\mathcal{E}_{el}=\frac{\lambda}{2}\varepsilon_{ii}^{2}+\mu\varepsilon_{ij}^{2}\ ,\quad\quad\varepsilon_{ij}=\frac{1}{2}\left(\partial_{j}u_{i}+\partial_{i}u_{j}\right)\ .\end{eqnarray}
 $\lambda,\,\mu$ are the Lamé constants. Two of the phenomenological
quantities introduced by phase field models appear here: $\kappa$,
a coefficient with dimensions of energy per unit length and
$g(\phi)$, a monotonic function that quantifies the degree
of\emph{ strain softening} due to material degradation. $g(\phi)$
satisfies $g(0)\!=\!0$, $g(1)\!=\!1$, $g'(0)\!=\! g'(1)\!=\!0$,
and $\lim_{\phi\to0}g(\phi)\!\sim\!\phi^{\alpha}$, with
$\alpha\!>\!2$. The latter condition was shown to ensure the
existence of crack-like solutions \cite{Karma.01}. We note that
$\phi$ in Eq. \ref{potentialE} does not break the rotational
invariance near the crack tip, unlike lattice or cohesive zone
approaches that inevitably introduce some near-tip anisotropy.
This anisotropy is questionable when amorphous materials, which
account for a large part of the available experimental data on
rapid fracture \cite{Fineberg.99}, are considered. Note also that
Eq. \ref{potentialE} is symmetric with respect to tension and
compression, though it is obvious that fracture occurs locally
under tension. This issue was addressed in \cite{Henry2004} by
introducing an ad-hoc asymmetry between tension and compression.

The equations of motion for a phase field model are variationally derived from the total potential energy
$E\!=\!\int\mathcal{E}\, dV$ and read \cite{Karma2004,Henry2004}
\begin{eqnarray}
\rho\partial_{tt}u_{i} & = & \partial_{j}\frac{\partial\mathcal{E}}{\partial H_{ij}}-\frac{\partial\mathcal{E}}{\partial u_{i}},\quad H_{ij}\equiv\partial_{j}u_{i}\ ,\label{EOM1}\\
\chi^{-1}\partial_{t}\phi & = & \partial_{j}\frac{\partial\mathcal{E}}{\partial\zeta_{j}}-\frac{\partial\mathcal{E}}{\partial\phi},\quad\zeta_{j}\equiv\partial_{j}\phi\ ,\label{EOM2}\end{eqnarray}
 where $\rho$ is the mass density. Translational invariance implies
that $\mathcal{E}$ is independent of $\mathbf{u}$, i.e. that $
\frac{\partial\mathcal{E}}{\partial u_{i}}=0$ .
 Noting that
$\frac{\partial\mathcal{E}}{\partial H_{ij}}=\sigma_{ij}$,
 Eq. \ref{EOM1} is, in fact, the usual linear momentum balance
equation, $\rho\partial_{tt}u_{i}=\partial_{j}\sigma_{ij}$.

The dynamics of $\phi$ in Eq. \ref{EOM2} are assumed to be
dissipative and hence first order in time. The last
phenomenological parameter $\chi$, needed to close the theory, has
the dimensions of volume per unit energy per unit time. Equations
(\ref{EOM1})-(\ref{EOM2}) are four coupled nonlinear equations
that can be solved numerically when the external boundary
conditions are specified. The simple choice of
$g(\phi)\!=\!4\phi^{3}\!-\!3\phi^{4}$, which satisfies all the
required conditions mentioned above, is commonly used in the
literature \cite{Karma.01,Karma2004,Henry2004}. It is important to
note that the usual traction-free boundary conditions on the crack
faces \cite{Freund.90} are not needed here since these are
automatically satisfied due to strain softening when $\phi\to0$.
Moreover, fracture in the phase field models is described by a
diffuse interface, which is a significant technical advantage,
since one does not have to track a moving boundary, which is
computationally expensive. The crack faces in the phase field
models are rather arbitrarily defined as a contour of fixed
$\phi$, usually $\phi\!=\!1/2$
\cite{Karma.01,Karma2004,Henry2004}.

\begin{figure}
\begin{centering}
\includegraphics[width=0.5\textwidth]{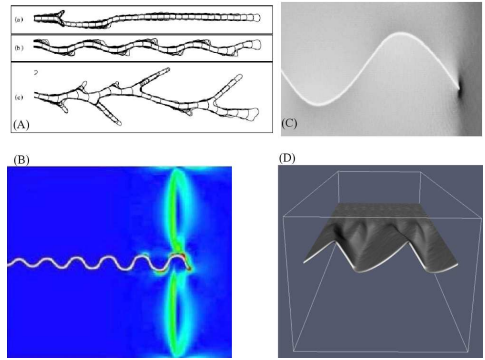}
\par\end{centering}

\caption{(A) Branching patterns in a two-dimensional out-of-plane
(Mode III) phase field model simulation \cite{Karma2004}. The
external loading level is increased from top to bottom. (B) Fast
sinusoidal crack tip oscillations in a two-dimensional in-plane
(Mode I) phase field model simulation under bi-axial loading
\cite{Henry2004}. (C) Quasistatic nonlinear crack tip oscillations
in a two-dimensional in-plane (Mode I) phase field model
simulation under thermal loading \cite{Corson.09}. (D) Complex
crack front morphology where crack breaks into segments in a
three-dimensional phase field model simulation under mixed mode
loading (in and out-of-plane, Modes
I+III), courtesy of A. Karma and A. J. Pons.\label{fig:phase-field}}

\end{figure}


A simple dimensional analysis, based on the phenomenologically
introduced parameters $\kappa$, $\mathcal{E}_{c}$ and $\chi$,
shows that the size of the region where LEFM breaks down, the
process zone, scales as \begin{equation}
\xi\sim\sqrt{\frac{\kappa}{\mathcal{E}_{c}}}\ ,\end{equation}
 the fracture energy $\Gamma$ scales as \begin{equation}
\Gamma\sim\sqrt{\kappa\,\mathcal{E}_{c}}\ ,\end{equation}
 and the timescale for energy dissipation near the tip of the crack
scales as \begin{equation}
\tau\sim\frac{1}{\mu\chi}\ .\end{equation}
 Therefore, in light of the discussion at the beginning of this Section,
the phase field models include a regularizing cut--off
length-scale $\xi$ and near-tip dissipation, resulting in a
well-defined and self-consistent mathematical description of a
fracture problem. Thus, crack tip shape, path and velocity
selection are expected to emerge naturally from the numerical
solution of Eqs. \ref{EOM1}-\ref{EOM2} in this class of
models. We note that the variational structure of the phase field
models allows a natural generalization of common energy-momentum
tensors and configurational forces in fracture mechanics
\cite{Hakim.09}.

Several groups have explored numerical solutions of the phase
field equations in various configurations. The results exhibit a
rather rich phenomenology that is at least in qualitative
agreement with experimentally observed phenomena; these include
two-dimensional steady-state crack motion in a strip geometry
above the Griffith threshold \cite{Karma.01,Spatschek.06},
branching instabilities \cite{Aranson.00,Karma2004,Henry.08},
rapid \cite{Henry2004} and quasi-static \cite{Corson.09}
oscillations and complex three-dimensional crack front dynamics. In Fig . \ref{fig:phase-field} we summarize some of
these results.

The phase field approach seems a very good mathematical method for
solving quasi-static fracture problems that are completely
controlled by LEFM and by geometric length-scales, cf.
\cite{Yuse.93}. In these cases it provides a proper crack tip
regularization and dissipation, and offers a computationally
efficient way to track complex crack configurations. For example,
the morphology of a quasi-static crack propagating under thermal
stresses was satisfactorily described in the framework of a phase
field model \cite{Corson.09}. In this example, all the pertinent
length-scales were geometric in nature and the crack path was
accurately described by the principle of local symmetry. On the
other hand, in situations where the crack tip physics has explicit
macroscopic manifestations
\cite{Sharon.2.96,Livne.07,Bouchbinder.07}, a number of important
open questions remains. The phase field methodology does\emph{
not} offer a physically realistic description of the deformation
and dissipative processes near crack tips. It is therefore
difficult to directly relate the phase field $\phi(\mathbf{x},t)$
itself or the phenomenological parameters $\kappa$ and $\chi$ to
experimental data. Therefore, this approach cannot, at present,
offer testable predictions in these situations. By coupling
directly to linear elasticity, it also does not incorporate
elastic nonlinearities (hence does not conform with Eq.
\ref{plastic_scale}). Future work is needed to further elucidate
the relation between the phase field models and the physics of
fracture. For example, incorporating nonlinear elastic effects as
discussed in Sect. \ref{sub:Small-Scales-Dominated} may be an
interesting future line of investigation.

\subsection{\label{sub:Atomic-Resolution}Atomic Resolution of Small Scales}

In some materials, details of material structure down to the
atomic scale play a decisive role in determining how details of
the fracture process work out. This statement is most evidently
true in crystals, which have natural cleavage planes. No matter
how big a sample of mica may be, it cracks easily along one easy
plane, and resists fracture along other directions
\cite{Obreimoff.30}. One way to characterize the effects of the
atomic scale is to say that $\Gamma$ depends upon orientation
relative to crystal planes, with cusp-like minima in special
directions. The surface energy of crystals depends upon
orientation in precisely this way \cite{Rottman.84}, and fracture
energy must have the same general behavior. The fracture of
crystals can be strongly anisotropic, even in a material that at
large scales has completely isotropic elastic behavior
\cite{Marder.IJF.04}.

While fracture energy has the same qualitative behavior as surface
energy in a crystal, the two are not equal. Fracture energy cannot
be less than surface energy because by definition surface energy
describes the minimum free energy needed to induce a solid to
separate along some plane. However it can be greater and in
general it is. When cracks travel through a crystal, the motion
through the periodic unit cells excites phonons. In general, any
object moving at velocity $v$ through a crystal with phonon
dispersion relation $\omega(k)$ excites phonons that obey the
Cherenkov condition \cite{Marder.00.CMP} $\omega(k)=vk$, and
cracks are no exception. While this fact has been confirmed
repeatedly in theoretical and numerical investigations, we are
unaware of any direct experimental evidence.

\begin{figure}
\begin{centering}
\includegraphics[width=.9\textwidth]{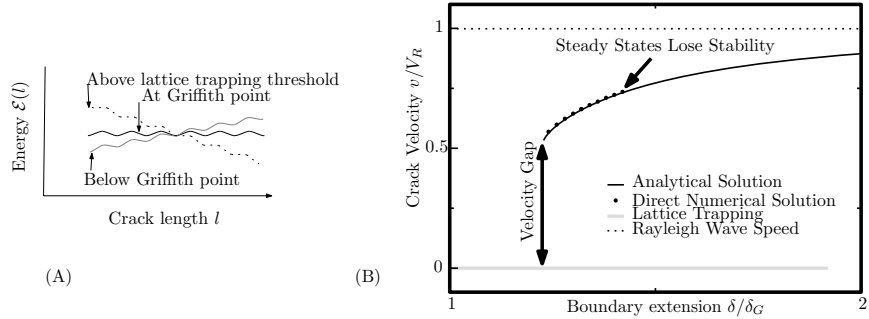}
\par\end{centering}

\caption{(A) Schematic view of how the energy of a crystal varies
as a function of the location $l$ of a crack tip within a fracture
plane. The potential energy surface is corrugated, and therefore
the crack cannot travel spontaneously at the Griffith point when
the elastic energy recovered per length equals the surface energy
cost. (B) Exact solution for crack velocity measured relative to
wave speed $V_{R}$ versus loading $\delta$ over Griffith loading
$\delta_{G}$ in a two-dimensional crystal at zero temperature,
showing lattice trapping, velocity gap, and the point where steady
states become dynamically unstable. Past the point of instability
steady states are technically impossible, but the deviations are
on an atomic scale that would be hard to detect
experimentally.\label{fig:Crystal1}}

\end{figure}

The obvious way to evade exciting phonons so that fracture and
surface energy coincide is for the crack to move very slowly.
Considering this possibility raises a number of interesting
problems. Thomson, Hsieh, and Rana \cite{Thomson.71} showed that
if one takes a crystal at zero temperature and very slowly pulls
its edges apart, when the stored energy reaches the Griffith
point, the crystal does not break. The strained crystal is
meta-stable. To see why, imagine that a crack moves along a plane
in a strip loaded as in Fig. \ref{fig:setup}B.  Let
$\mathcal{E}(l)$ give the minimum-energy atomic configuration
subject to the constraint that the crack tip be at position $l$.
The crack location is easy to define as a continuously varying
function of atomic positions. Finding the function
$\mathcal{E}(l)$ in practice is demanding
\cite{Henkelman.00,E.02}, but in principle one can see immediately
what shape it has to have, as shown in Fig. \ref{fig:Crystal1}.
First $\mathcal{E}(l)$ must vary linearly with $l$, with positive
slope when $\delta$ is below the Griffith point and negative slope
when it is stretched above the Griffith point. In addition,
$\mathcal{E}(l)$ must have a periodic component corresponding to
different locations in the unit cell. Right at the Griffith point
the crack tip sees a corrugated potential, is necessarily trapped
in a local minimum, and does not propagate. The crystal must be
stretched by an extra amount so that the slope of $\mathcal{E}(l)$
is negative everywhere. Thomson, Hsieh, and Rana \cite{Thomson.71}
called this phenomenon lattice trapping.

This discussion of linear stability does not immediately rule out
the possibility that a crack could move very slowly just above the
Griffith threshold if it could only get started. Exact solutions
for cracks running in lattices found by Slepyan
\cite{Slepyan.81,Slepyan.02} and Marder \cite{Marder.95.jmps} show
that this cannot happen either. Every time a crack moves to a new
unit cell it slides down the corrugated potential and causes the
atoms there to vibrate at frequencies on the order of $c/a$ where
$c$ is a sound speed and $a$ is a lattice spacing. If the atom
vibrates many times it radiates energy which is lost to crack
motion. Therefore the most energy-efficient crack motion requires
the crack to get to a new cell during a time on the order of a
vibrational period. This means that its velocity must be a
fraction of the sound speed, and if the velocity drops below a
threshold it cannot move at all, as shown in Fig.
\ref{fig:Crystal1}B. Thus there is a finite velocity gap for
cracks moving in crystals at zero temperature. Either cracks move
at a finite fraction of the sound speed and spend energy on
phonons or they do not move at all.

All of these conclusions must be modified at nonzero temperature.
If the energy of thermal fluctuations is comparable to the size of
the corrugations in Fig. \ref{fig:Crystal1}A, then the
corrugations should be washed out and the crack should behave as
one would expect if the material is continuous down to the finest
scales.

\begin{figure}
\begin{centering}
\includegraphics[width=0.9\textwidth]{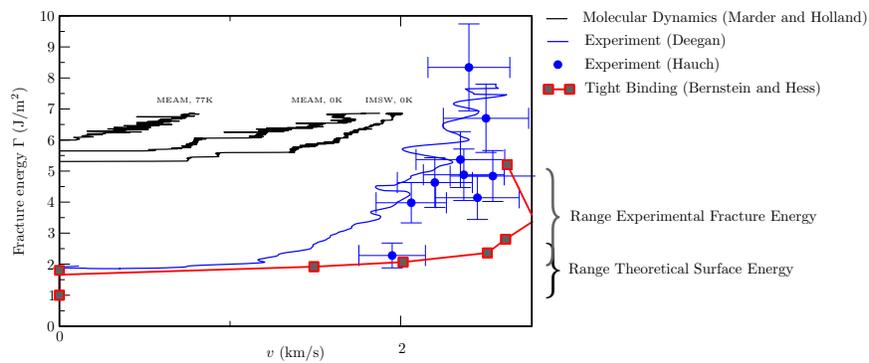}
\par\end{centering}

\caption{Comparison of experiment and theory for the fracture of
silicon along the (111) plane, showing the energy $\Gamma(v)$
needed for propagation at various speeds. The experiments were
carried out at room temperature in silicon single crystals
\cite{Hauch.99}. Deegan established an energy gradient in a strip
so that the curve could be mapped out in a single run. The three
upper theoretical curves are molecular dynamics simulations using
the Modified Embedded Atom Method (MEAM) at temperatures of 0 and
77K, and the Inadvertently Modified Stillinger Weber (IMSW)
potential \cite{Holland.99}. Lattice trapping is so large for
these potentials that they disagree substantially with experiment.
However a tight-binding computation of Bernstein and Hess is in
agreement \cite{Bernstein.03}.\label{fig:Silicon}}

\end{figure}

Attempts to follow through on all these ideas and compare with
experiment has proceeded the furthest in single-crystal silicon,
where the situation is not completely settled. The minimum energy
plane in silicon is (111), and theoretical estimates for the
surface energy $2\gamma$ range from 1 to 2.5 J/m$^{2}$
\cite{Spence.93}. Experimental measurements of the surface energy
are in the range of 2-5 J/m$^{2}$ \cite{Spence.93}, but as
dynamically propagating cracks have been observed with energies as
low as 2.25 $\pm0.25$ J/m$^{2}$ \cite{Hauch.99}, the lower end of
the range is most likely. Both experimental and numerical
estimates of $\Gamma(v)$ are available, as shown in Fig.
\ref{fig:Silicon}. One might expect based upon Fig.
\ref{fig:Crystal1}A that the detailed relationship between
energy available for fracture $\Gamma$ and crack speed $v$ might
depend rather sensitively upon details of the potential energy
surface, particularly upon the height of the corrugations, and
this is the case. The mechanical behavior of silicon at the atomic
scale is often described by classical empirical potentials
\cite{Holland.99}, such as the one due to Stillinger and Weber
\cite{Stillinger.85}, or the Modified Embedded Atom Method
\cite{Baskes.92}. All of these empirical potentials appear to have
corrugations that are too large, and therefore produce
substantially more lattice trapping and larger velocity gaps than
are seen in experiment. The potentials are constrained by
considerations of crystalline symmetry, fracture energy, and
low-energy properties like sound speed. No one has yet managed to
guess potentials for silicon that gets all these things right and
the fracture properties as well, although the Modified Embedded
Atom Method is so far the best. However, an elaborate calculation
of Bernstein and Hess \cite{Bernstein.03} that couples
tight-binding quantum mechanics near the crack tip to empirical
potentials further away does seem to capture all of the observed
experimental features well. It is a bit surprising that such
effort must be expended to get the energy at which fracture begins
right within a factor of 2.

\subsection{\label{sub:Supersonic-States}Supersonic States: a New Energy Scale}

The preceding sections showed some effects of integrating length
scales with LEFM. An additional assumption of LEFM (cf. Sect.
\ref{sub:Linear-Elastic-Fracture}) was that energy must flow into
a crack's tip to enable it to propagate. Here we will examine the
consequences of discarding this assumption.

The Rayleigh wave speed as the upper limit for a crack was a
consequence of this assumption. One can ask what limits apply to
crack speeds if both the tendency to instability is suppressed and
the above assumption discarded. Suppressing instabilities can be
arranged by having the fracture move along a weak interface, or by
finding a material which for some reason does not permit crack
branching to occur. Cracks travelling along weak interfaces have
been of greatest interest in the study of earthquakes. Andrews
\cite{Andrews.76} showed numerically that shear cracks on a weak
interface could travel faster than the shear wave speed. This
problem was studied further by Burridge, Conn, and Freund
\cite{Burridge.79}, who showed that LEFM permits shear cracks to
travel precisely at a speed given by $\sqrt{2}$ times the shear
wave speed, but that the range of possible speeds is broadened by
cohesive forces near the crack tip.

Since we have maintained that LEFM must be supplemented with
information from smaller scales in order to make predictions about
crack velocities, we should explain how this theory has been
employed to make predictions about the limiting speeds of cracks.
One can use LEFM to compute the energy travelling from far elastic
fields in to the crack tip assuming that the crack is travelling
at velocity $v.$ For velocities $v$ lower than the Rayleigh wave
speed, this energy flux is always positive. For velocities above
the Rayleigh wave speed, however, it is either imaginary or
negative. There is one special case, of shear cracks travelling at
$\sqrt{2}$ times the shear wave speed, where the energy flux just
vanishes, and this is the physically allowed value. Cracks in
tension, by this logic, are forbidden to travel faster than the
shear wave speed.

\begin{figure}
\begin{centering}
\includegraphics[width=0.9\textwidth]{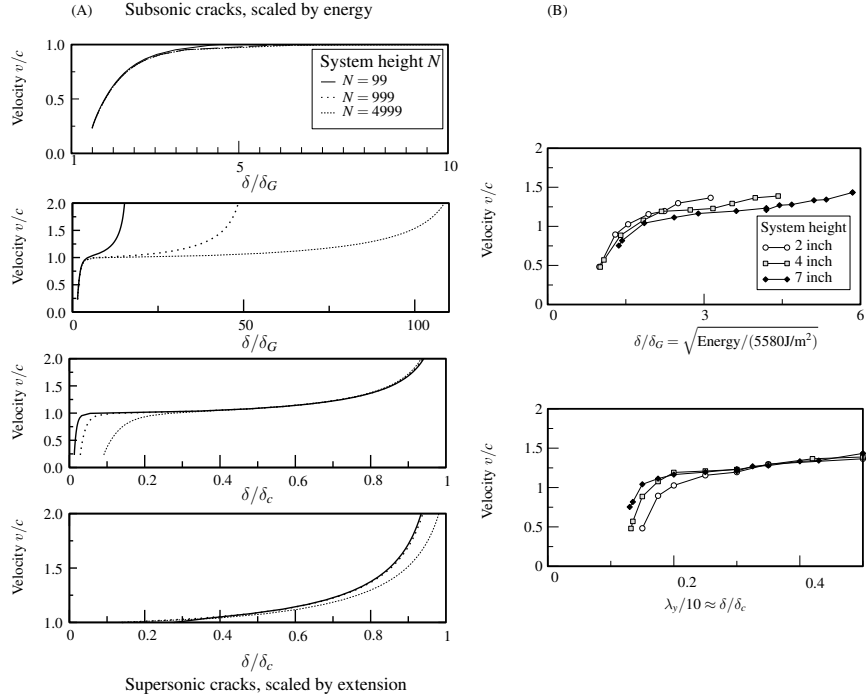}
\par\end{centering}

\caption{(A) Four different views of Neo--Hookean crack
velocities, showing that, depending upon how they are scaled and
displayed, one focuses either upon conventional subsonic fractures
or supersonic ruptures. In the limit of infinite system height
$N$, the two different types of solutions are separated by an
infinitely long plateau at the wave speed. The final panel simply
re-plots the data of the third panel with a different range of
velocities visible, emphasizing that if the macroscopic limit is
taken holding strain constant, it appears that all cracks are
supersonic. (B) Experiments in rubber at 85 C. The temperature
needs to be elevated because at room temperature rubber undergoes
a transition above extension ratios of 3.5 where its fracture
energy increases enormously. The experiments are carried out in
strip geometries with four different system heights. Velocities
are scaled by the wave speed $c=\mbox{21.9}$ m/s. The
dimensionless extension $\delta/\delta_{G}$ is estimated by the
square root of the energy scaled by the Griffith energy of 5580
J/m$^{2}$. The vertical extension ratio $\lambda_{y}$, the ratio
of the stretched height to the original height, is scaled by 10,
which is a rough estimate of the critical extension
$\delta/\delta_{c}$ obtained by comparing slopes in right and left
panels. Experimental results courtesy of Hepeng Zhang, Johnathan
Niemczura, and K. Ravi-Chandar.\label{cap:Four-different-views}}

\end{figure}

Nevertheless, Buehler, Abraham, and Gao \cite{Buehler.03} observed
such cracks in molecular dynamics simulations, and Petersan
\emph{et al.} \cite{Petersan.04} observed them in experiments in
rubber.  They
evade the apparent limitations of linear elastic theory because
they correspond to a completely different scaling theory.

In LEFM, the scaling parameter is $\delta_{G}$, the total displacement
of the boundaries at which the energy reaches the Griffith threshold
and there is enough elastic potential energy available per unit length
to create new surface. There is another critical extension, however,
$\delta_{c}$, which is the extension at which the solid would be
stretched so much that the bonds between adjacent material points
would give way completely. For most brittle materials, reaching $\delta_{c}$
requires stretching the solid to around 20\% more than its original
length. In rubber, it requires stretching the solid to 8 or 10 times
its original length.

If a crack tip avoids going unstable, then eventually when the
boundary extension $\delta$ reaches a fraction of $\delta_{c}$ the
crack speed becomes supersonic. The signature of the supersonic
states lies in their behavior as the size of the system increases.
For conventional subsonic cracks, cracks in two systems behave
in the same way when $\delta_{G}$ is the same. Thus if one strip
is twice the height of another, cracks will have the same velocity
in the two strips if the second is stretched by a factor of
$\sqrt{2}$ times as much as the first. For supersonic cracks,
strain itself is the controlling variable, and energies rise to a new
scale. Thus in the supersonic
case cracks in the second strip will look like those in the first
when it is stretched to a height $\delta$ twice as much. Fig.
\ref{cap:Four-different-views} shows both both analytical
calculations and experiments in rubber where these scaling
behaviors are observed. The analytical calculations are carried
out in exactly solvable lattice models \cite{Marder.JMPS.2006} and
the experiments are carried out in rubber. The calculations in the
figure are for a simple case of anti-plane shear (Mode III), but they have
been extended to fracture in tension as well \cite{Guozden.09}. Rubber
may seem an odd experimental setting for fracture mechanics, but the
relation between force and displacement is nearly linear over wide
range of displacements, and the strongest deviations from linear
behavior are confined to a small region near the tip as required by
small-scale yielding.

The fundamental reason that cracks normally travel slower than all
wave speeds is that they are only able to propagate by absorbing
all the potential energy stored in the sample, all the way out to
the boundaries. In the case of supersonic cracks, the solid has
been stretched so much that energy sufficient to snap bonds is
located with a fixed distance of the tip. Thus the crack speed is
no longer limited by the time needed to transport energy to the
tip from far away. Continuum elastic solutions for supersonic
cracks show that their tip must now be a \emph{source} of energy,
a fact that has been used to argue that they cannot exist. In
fact, enough energy is stored near the tip of a supersonic crack
that it can break bonds with enough left over to travel outwards.
A final observation is that in samples strained enough to support
supersonic cracks, most of the energy in the sample is released
through contraction of material after the crack has passed. This
energy must be absorbed by some mechanism; in rubber, it is
absorbed by bulk dissipation, which is particularly strong along a
shock line that develops in the wake of the crack.

\section{Conclusions and Unsolved Problems}

We took a careful look at the underlying assumptions that describe
our current view (LEFM) of how the simplest cracks evolve, and
noted a few possible ``cracks'' in its underlying structure.
\begin{itemize}
\item LEFM is a scale-free theory, but a fundamental understanding of the
physics of fracture requires description at the scales where
rupture processes occur.
\item Fracture processes are usually not explained by thermodynamic ground
state arguments, and they are not even necessarily controlled by flow
of energy to the crack tip.
\item The major unsolved problems of fracture concern the way the crack
line decides how and where to move. Problems include the reasons cracks
change direction while in motion, and conditions for crack fronts
to remain essentially planar or to create complex ramified surfaces.
\end{itemize}
The resolution of these issues may be intimately related to both
our understanding of the proper ``ground state'' of this system
and its stability. We have seen that there may not be a single
solution to this problem, and different classes of physically
viable solutions may exist. We surmise that a fundamental
description of the stability of such solutions for ``simple''
cracks could be closely linked to how to properly take into
account the scales that define them. These scales may be dynamic
entities, as we have seen, in sections
\ref{sub:Small-Scales-Dominated} and \ref{sub:Phase-Field-Models}
or static ones, as in sections \ref{sub:Atomic-Resolution} and
\ref{sub:Supersonic-States}. They may simply determine the
character of the ground state, or possess, themselves, intrinsic
complex dynamics that may undermine ground-state stability. These
issues define current active research directions. Perhaps their
resolution (in the coming decade) will provide the basis of our
next review.

\bibliographystyle{ieeetr}
\bibliography{fracture}

\end{document}